\documentclass{epl}
\usepackage{graphics}
\usepackage{latexsym,amssymb}
\title{Kink instability of a highly deformable elastic cylinder}
\author{Apurba Lal Das\inst{1} \and Animangsu Ghatak\inst{1}}

\institute{ \inst{1} Department of Chemical Engineering, Indian
Institute of Technology, Kanpur 208016, India }

\begin{document}

\maketitle

\begin{abstract}
When a soft elastic cylinder is bent beyond a critical radius of
curvature, a sharp fold in the form of a kink appears at its inner
side while the outer side remains smooth. The critical radius
increases linearly with the diameter of the cylinder while
remaining independent of its elastic modulus, although, its
maximum deflection at the location of the kink depends on both the
diameter and the modulus of the cylinders. Experiments are done
also with annular cylinders of varying wall thickness which
exhibits both the kinking and the ovalization of the
cross-section. The kinking phenomenon appears to occur by extreme
localization of curvature at the inner side of a post-buckled
cylinder.
\end{abstract}

%\paragraph{\textbf{Introduction}}

Highly deformable, soft elastic materials occur in many different
applications e.g. soft tissues, artificial organs, therapeutic
patches, shock absorbers, dampeners, platforms for micro-fluidic
devices etc. In these variety of applications the material is
exposed to many different form of mechanical loads, which, due to
the large deformability of these materials, can generate such
responses which are different from that commonly observed with the
liner elastic systems. An example is the surface wrinkling which is
the most common form of mechanical response in elastic objects
subjected to compressive stresses e.g. engendered by bending of an
elastic block \cite{Gent99}, or by sliding of a rubber block over a
hard surface \cite{Schallamach71} or by the transverse poisson
contraction of a uniaxially stretched rubber sheet \cite{Cerda2003};
however, here we report a different kind of behavior observed with
soft hydrogel under similar circumstances. Our experiments with
hydrogel cylinders bent beyond a critical curvature show that in
stead of wrinkling, the material here responds by the appearance of
one single sharp fold in the form of a kink at the inner side of the
cylinder. As the cylinder is progressively bent, at a critical
radius of curvature, the kink appears with a abrupt jump accompanied
by the reversal of the curvature in the region close proximity to
the kink; at the kink the curvature shoots up to infinity. While
kinking phenomenon has been observed with slender biological
filaments like DNA \cite{Koehler2000} or bacterial flagella
\cite{Shaevitz2005} and inorganic fibers like multi-walled
nano-tubes \cite{Cohen2003}, where kinks appear rather
intrinsically, mediated by the dual effects of local defects and
intermolecular interactions, the kinking phenomenon observed in our
system appears to be akin to the classical Euler's buckling
instability\cite{Love27}, albeit localized at the inner side of a
post-buckled cylinder. In this letter we have characterized this
instability by using solid and annular cylinders of varying inner
and outer diameters and by varying the modulus of the gel material.

\begin{figure}[!htbp]
\centering
\includegraphics[width=9.0cm]{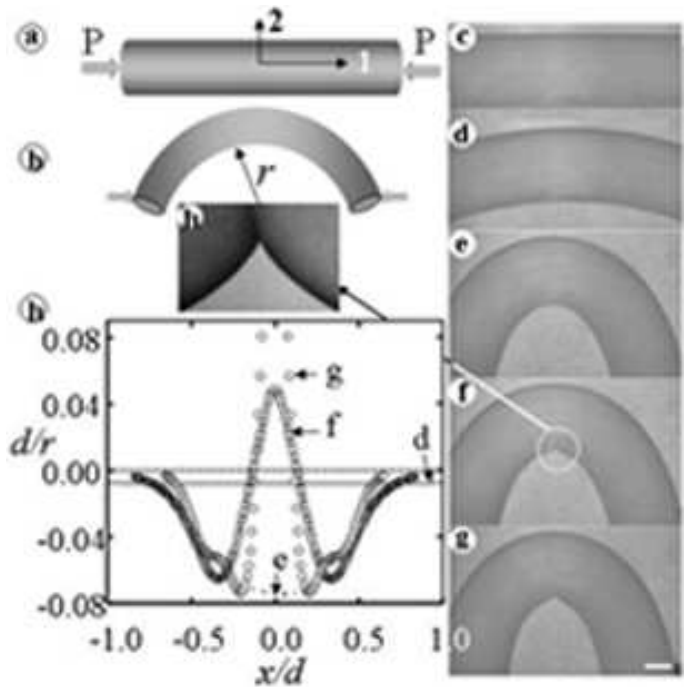}
\caption{(a) Schematic of the experiment, in which a straight gel
cylinder of circular or rectangular cross-section is subjected to
compressive end loads. (b) At a critical load, the cylinder buckles
(Euler's buckling instability \cite{Love27}), the curvature of which
increases with further increase in the load. (c)-(g) A typical
sequence of video-micrographs (captured using a video-camera having
25 fps resolution) leading to the appearance of the kink in a
cylindrical gel cylinder of shear modulus 48 kPa and diameter
$d=27$mm. While (c) and (d) correspond to bending of the cylinder
much before the appearance of the kink, micrograph (e) appears just
before the occurrence of the kink in (f) and (g) corresponds to
bending long after (f). Videomicrograph (h) depicts a magnified
picture of the cylinder closed to the kink. (i) The curvature at the
inner side of the cylinder in micrographs (d) to (g) are
nondimensionalized as $K = d/r$ and plotted with respect to its
dimensionless length $L=x/d$ along direction 1.} \label{fig:fig1}
\end{figure}

These gel samples were prepared by polymerization in water of
Acrylamide as the monomer with the NN'-Methylene bis-Acrylamide as
the cross-linker ($0.267$\% by weight of the monomer), TEMED as the
promotor ($0.004$\%) and Ammonium per sulfate ($0.04$\% by weight of
the monomer) as the initiator. The monomer to water ratio was varied
in such a way that the final product contained $60-95$\% by weight
of water. The polymerization reaction was carried out in molds of
different sizes to prepare samples of variety of thicknesses and
diameters. The gel samples were washed in water for at least eight
hours in order to remove any un-reacted monomer and cross-linker and
were subjected to the experiment depicted in figure \ref{fig:fig1}.
Since, in the time scale of our experiment ($\sim 1$ min) the gel
samples behaved like an isotropic elastic solid, these were
characterized by their shear modulus $\mu$. The modulus was
estimated by cantilever beam experiment \cite{Landau} in which one
end of the cylinder was clamped to a rigid support while the
deflection of the free end under gravity was measured. Small enough
length of the cylinder ensured that it did not stretch under
gravity. We carried out also compression test \cite{Muniz2001} on
the gel samples which yielded similar values for the modulus.
Systematic variation of initial water content in the pre-polymer
solution allowed us to produce gels of different modulus.

Figure \ref{fig:fig1} (a) and (b) depict the schematic of the
experiment, in which cylinders of varying diameter ($5-50$ mm) and
shear modulus ($0.5-50$ kPa) were subjected to compressive end
loads between two rigid supports. Since very low modulus gels were
observed to undergo deformation under their own weight, these gel
samples were first immersed in water in which the samples remain
neutrally buoyant and were then subjected to the experiment of
figure \ref{fig:fig1}. In such experiments the straight cylinder
did not remain stable but at a critical load underwent buckling;
this phenomenon is the well-known Euler's buckling instability
which resulted in smooth bending at both the inner and the outer
sides of the rod; however, when the curvature at the inner side
exceeded a critical value $r_c$, a sharp fold appeared in the form
of a kink. Video-images \ref{fig:fig1}(c)-(f) represent a typical
sequence as a cylinder of diameter $d = 27$mm and shear modulus
$\mu = 31$ kPa was progressively bent leading up-to the appearance
of the kink. While the figure \ref{fig:fig1}(f) represents the
frame at which the kink just appeared, frames (d) and (e)
represent respectively a cylinder which was bent slightly and then
critically; frame \ref{fig:fig1}(g) depicts the bending of the
cylinder long after the appearance of the kink. The region closed
to the kink is magnified in figure \ref{fig:fig1}(h) which shows
that curvature changed from negative to positive at the vicinity
of the kink. This observation is quantified in figure
\ref{fig:fig1}(i) in which we plot the curvature of the inner side
of the cylinder as it appears in images of figure
\ref{fig:fig1}(d)-(g) as a function of distance along the axis of
the cylinder. The figure shows that for images (c) and (d), the
curvature is always negative through out the length of the
cylinder; however in figure \ref{fig:fig1}(e) the curvature
changes abruptly to positive at the vicinity of the kink. In fact,
at the kink, the curvature reaches infinity as the radius of
curvature becomes zero. Thus curve (e) represents the critical
bending after which curvature get localized within a distance
$\sim d$ from the location of the kink. Once localized, the kink
behaves like a hinge, so that, with further bending, the curvature
does not change any significantly as evident from curve (g).
Whereas all these experiments were done with temporal resolution
of $0.04$ sec, few experiments with higher resolution ($\sim
0.001$ sec) showed \cite{Movie} that the evolution of the kink was
complete within a timescale of $0.001-0.005$ second which was of
the same order as that of the elastic deformation of the cylinder:
$\tau_{b} \sim \delta \left(\rho/E\right)^{1/2} \sim 0.01-0.001$
sec ($\delta = 2$ mm, $\rho = 1.0$ gm/cc, $E=50$ kPa). In fact
this timescale was orders of magnitude smaller than the
poro-elastic timescale proposed by Skotheim et al
\cite{Skotheim2004,Forterre2005}: $\tau_p = 10^2-10^4$ sec
\cite{poroelastic}. Furthermore, in the reverse cycle, as the
bending load was immediately withdrawn, the kink disappeared
albeit at a slightly larger radius of curvature and the cylinder
straightened out. However, in long time scale the kinking
phenomenon was not completely reversible, because, when the
cylinder was kept in the state as in frame \ref{fig:fig1}f for a
prolonged period ($3-10$ min), a defect appeared at the location
of the kink, so that in subsequent cycles, the kink appeared at
the same location at a much lower curvature. While this
irreversibility is a signature of complex rheological character of
the gel, we will concern here about its short time response only
which is fairly reversible so that laws of elasticity are
applicable.

\begin{figure}[!htbp]
\centering
\includegraphics[width=7.0cm]{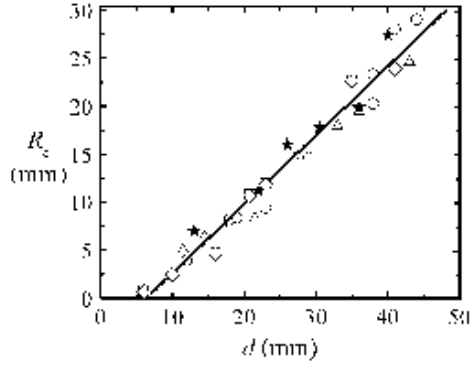}
\caption{Critical inner radius of curvature at which the kink
appears is plotted against the diameter of the cylinder. The symbols
$\diamondsuit $, $\circ $, $\bigstar$, $\Box$, $\Delta $ represent
the data from experiments with gels of shear modulus $8$, $14$,
$22$, $31$ and $45$ kPa respectively. Data from experiments with
cylinders of different modulus fall on a single straight line with
result $r_c = 0.72d - 4.54$.} \label{fig:fig3}
\end{figure}
We carried out experiments at fresh locations of the cylinders to
obtain the data of critical radius of curvature $r_c$. The data
for cylinders with varying diameter ($3-50$ mm) and modulus
($5-50$ kPa) as summarized in figure \ref{fig:fig3} shows that all
data fall on a single straight line with the result: $r_c = 0.72d
- 4.54$ implying the existence of a minimum diameter of the gel:
$d_{min} \sim 6.3$ mm, below which the kink did not appear. For
these cylinders $d<d_{min}$ mm, progressive bending resulted in
increase in curvature while its inner and outer side remained
smooth. For gel cylinders with $d>d_{min}$, the $r_c$ remains
independent of $\mu$. Interestingly, similar results were obtained
also by Gent et al \cite{Gent99} in experiments with rubber block
of shear modulus $\mu \sim 1.5-1.8$ MPa. Bending of rubber blocks
of thickness $h = 7-14$ mm resulted in surface wrinkling at the
inner side when the radius of curvature $r_c \leq 0.72h$. This
similarity signifies that the critical curvature at which surface
instabilities ensue remain unaltered over three orders of
magnitude in values of $\mu$, although, once ensued they evolve
differently leading to either the wrinkles or the kink.

\begin{figure}[!htbp]
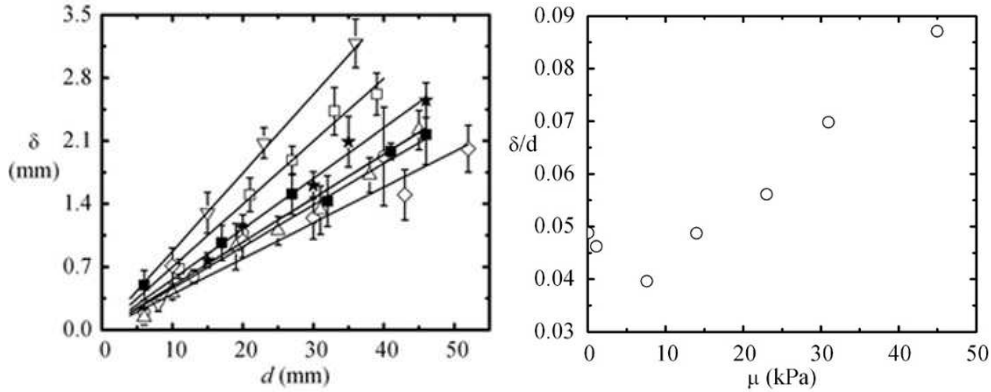

\centering
\includegraphics[width=6.5cm]{delta_vs_d21.eps}
\includegraphics[width=6.5cm]{deltavsmu1.eps}
\caption{(a) The displacement $\delta$ of the gel at the location of
appearance of the kink is plotted against the diameter $d$ of the
cylinder. The symbols $\nabla$, $\square$, $\bigstar$, $\circ$,
$\diamond $, $\bigtriangleup$, $\blacksquare$ represent gels of
modulus $45$, $31$, $22$, $14$, $8$, $1$ and $0.5$ respectively. (b)
The slope $\delta/d$ of these curves are plotted against the modulus
$\mu$ of the gel.} \label{fig:fig4}
\end{figure}
The displacement $\delta$ at the kink however varies with both the
diameter of the cylinders and their modulus. $\delta$ increases
linearly with the diameter but with $\mu$, it varies
non-monotonically as shown in figure \ref{fig:fig4}. $\delta/d$
decreases in the range $\mu =0.5-1$ kPa but increases when $\mu =
1-50$ kPa. This result shows that although geometry triggers the
instability, its evolution is mediated by the dual effects of both
the material and the geometric properties of the gel cylinders.

\begin{figure}[!htbp]
\centering
\includegraphics[width=8.0 cm]{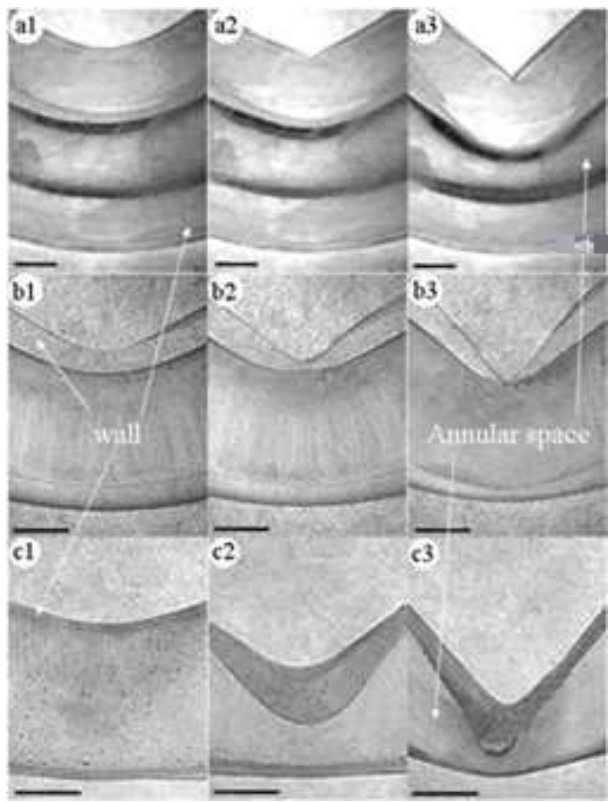}
\caption{Annular gel cylinders are subjected to experiment in figure
\ref{fig:fig1}. Video-images $a_i, b_i, c_i$ correspond to cylinders
of inner diameter $d_i = 26$ mm and wall thickness $t_i = 28, 8$ mm
and $4$ mm respectively. In each series $i = 1, 2$ and $3$
correspond to images in successive frames respectively. In images
$a_i$ we see the appearance of the kink followed by the ovalization
of the cross-section of the cylinder; in $b_i$ both kink and
ovalization occur simultaneously; whereas in $c_i$ we see only the
ovalization phenomenon.} \label{fig:fig2}
\end{figure}
In order to further probe the effect of geometry we performed
similar experiments with annular cylinders of which the thickness of
the wall was varied. While for the solid cylinders, appearance of
kink was the only mode of response, the annular ones responded by an
additional mode: by ovalization of the cross-section of the
cylinder, known as the Brazier effect \cite{Brazier26}. The images
in figure \ref{fig:fig2} represent the typical patterns which were
obtained when cylinders of inner diameter $d_i = 26$ mm and wall
thickness $t_i = 4-28$ mm were subjected to bending. Here we see the
appearance of not only the kinking instability as with solid
cylinders, but also ovalization of its cross-section. Both these
phenomena are captured in the sequence of images
\ref{fig:fig2}a$_{1-3}$ obtained while bending a cylinder of wall
thickness $t_i = 28$mm. As bending exceeded the critical curvature,
the kink appeared first at the inner side followed by the
ovalization of its cross-section. These two phenomena however
occurred simultaneously as the wall thickness decreased, as is
evident in the sequence \ref{fig:fig2}b$_{1-3}$ obtained with $t_i =
8$mm. Both the ovalization and the kinking instability are evident
in figure \ref{fig:fig2}b$_{2}$-b$_{3}$. When we further decreased
the thickness to $t_i = 4$mm, we observed only the Brazier effect
and not the kinking instability as in figure
\ref{fig:fig2}c$_{1-3}$.

We rationalize our observations by considering the deflections of
a rectangular elastic slab subjected to compressive stress
\cite{Biot}. A slab of thickness $h_0$ and length $L_0$ is
compressed by an end stress $P$ to length $L = \lambda_1 L_0$ and
thickness to $h = \lambda_2 h_0$ where $\lambda_1$ and $\lambda_2$
are the initial compression ratios along $x$ and $y$ directions
respectively; $\lambda_3 = 1/\lambda_1 \lambda_2$ is that along
$z$. The slab buckles over and above this initial deformation
following the stress equilibrium and incompressibility relations
depicted in terms of the incremental stresses $s_{11}$, $s_{12}$
and $s_{22}$ and displacements $u$ and $v$ along $x$ and $y$
respectively:
\begin{eqnarray}\label{eq:eq1}
{s_{11}}_x + {s_{12}}_y -Pw_y & = & {s_{12}}_x + {s_{22}}_y -Pw_x = 0 \nonumber \\
u_x + v_y & = & 0
\end{eqnarray}
The incremental quantities are related to the derivatives of
displacements $u$ and $v$ as,
\begin{eqnarray}\label{eq:eq2}
s_{11} - s & = & 2\mu u_x, \hspace{1mm} s_{22} - s = 2\mu v_y,
\hspace{1mm} s_{12} = \mu \left(u_y + v_x \right) \nonumber \\
\mu & = & \mu_0 \left({\lambda_1}^2 - {\lambda_2}^2\right)
\end{eqnarray}
where $s$ is the average stress: $s = \left(s_{11} +
s_{22}\right)/2$. Equations \ref{eq:eq1} and \ref{eq:eq2} are
solved using the boundary conditions that normal and shear
traction on the free surfaces are zero:
\begin{eqnarray}\label{eq:eq3}
\Delta f_x = s_{12} + Pe_{xy} = 0 = \Delta f_y = s_{22}
\end{eqnarray}
The above equations have non-trivial solutions for the
displacements, as long as the buckling load and the aspect ratio of
cylinder satisfy the following conditions \cite{Biot}:
\begin{eqnarray}\label{eq:eq7}
4k \tanh\left(k\gamma\right) - \left(1+k^2\right)^2
\tanh\left(\gamma\right) = 0 \nonumber \\
\gamma = \left(\pi h/L\right) \hspace{3mm} k =
\sqrt{\left(1-P/2\mu\right)/\left(1+P/2\mu\right)}
\end{eqnarray}
which in essence is a characteristic equation depicting the
critical condition for buckling of the cylinder. While in the
limit of a slender slab, $L/h >> 1$ i.e. when $\gamma \rightarrow
0$, \ref{eq:eq7} yields the Euler's buckling load, in the other
extreme of a thick semi-infinite slab i.e. when $L/h << 1$ i.e.
$\gamma \rightarrow \infty$, it leads to a critical curvature at
the inner side at which surface wrinkles appear. Gent et al
\cite{Gent99} calculated this critical radius: $r_c/h_0 \approx
0.36$ which is about half of that obtained in experiments. This
discrepancy is explained if we consider that wrinkling is a local
phenomenon and occurs over and above the buckling instability. In
fact figure \ref{fig:fig1}i suggests that the instability occurs
within a distance $L \approx h$, so that $\gamma = \pi$ and
solving equation \ref{eq:eq7}, $k$ is obtained as $0.4$. Following
Gent et al's calculation \cite{Gent99}, the critical radius of
curvature is then obtained as $r_c/h_0 \approx 0.6$ which is
rather close to what was seen in experiments. Although this
analysis captures the critical condition for occurrence of the
instability, transient analysis of the system is needed in order
to understand the evolution of the instability as well as
displacement $\delta$ at the location of the kink.

To summarize, we have described here a new mode of surface
undulation which results in a kink at the compressed side of a bent
hydrogel cylinder. The timescale of evolution of this instability
compares well with that of the elastic deformation of the material
which implies that it is an elastic response of the material. The
critical curvature at which the instability appeared could be
captured by a simple approximate analysis but detail non-linear
theory is needed to understand the evolution of the kink.

\acknowledgements This work was supported by the Research initiation
grant of IIT Kanpur.

$^*$ Electronic address: aghatak@iitk.ac.in


\begin{thebibliography}{200}
%
\bibitem{Gent99} A. N. Gent and I. S. Cho, Rubber Chem. Technol. {\bf
72}, 253(1999).
%
\bibitem{Schallamach71} A. Schallamach, Wear {\bf 17}, 301(1971).
%
\bibitem{Cerda2003} E. Cerda and L. Mahadevan, Phys. Rev. Lett. {\bf 90},
074302(2003).
%
\bibitem{Koehler2000} S. A. Koehler, T. R. Powers, Phys. Rev. Lett. {\bf 85},
4827(2000)
%
\bibitem{Shaevitz2005} J. W. Shaevitz, J. Y. Lee2 and D. A.
Fletcher, Cell {\bf 122}, 941(2005).
%
\bibitem{Cohen2003} A. E. Cohen and L. Mahadevan, Proc. Natl. Accad. Sci. {\bf 100}, 12141(2003).
%
\bibitem{Love27} A. E. H. Love, "A treatise on the mathemetical
theory of elasticity", Dover publications, New York 1944.
%
\bibitem{Landau} L. D. Landau and E. M. Lifshitz, "Theory of elasticity 3rd ed.", Butterworth-Heienemann, Oxford 2000.
%
\bibitem{Muniz2001} E. C. Muniz and G. Geuskens, Macromolecules,
{\bf 34}, 4480(2001).
%
\bibitem{Movie} A typical movie of the kinking instability can be
seen in "http://home.iitk.ac.in/$\sim$aghatak/Moviepage.htm"
%
\bibitem{Skotheim2004} J. M. Skotheim and L. Mahadevan, Proc. R. Soc. Lond. A, {\bf 460},
1995(2004).
%
\bibitem{Forterre2005} Y. Forterre, J. M. Skotheim, J. Dumais and L. Mahadevan, Nature
{\bf 433}, 421 (2005).
%
\bibitem{poroelastic} Poroelastic timescale $\tau_p = \mu \delta^2/kB$ in
which $k$ is the hydraulic permeability of the gel, $\mu$ is the
viscosity of water inside the network and $B$ is the bulk modulus of
the dry gel. The quantity $kB/\mu$ is a diffusion constant
\cite{Tanaka} which for our hydrogel samples should be $\approx
10^{-8} - 10^{-10}$m$^2$/sec. In our experiments, $\delta = 1-3$mm
so that $\tau_p = 100-10^4$ sec.
%
\bibitem{Tanaka} T. Tanaka, L. O. Hocker and G. B. Benedek, J.
Chem. Phys. {\bf 59}, 5151(1973).
%
\bibitem{Brazier26} L. G. Brazier, Proc. Roy. Soc. Lond. A CXVI  104(1926).
%
\bibitem{Biot} M. Biot, "Theory of incramental deformations", Dover publications, New York 1944.

\end{thebibliography}
\end{document}